# Bin Packing/Covering with Delivery: Some variations, theoretical results and efficient offline algorithms

György Dósa[*]    Zsolt Tuza[†]

Latest update on 2011-2-11

## Abstract

In the recent paper [2] we introduced a new problem that we call Bin Packing/Covering with Delivery, or BP/CD for short. Mainly we mean under this expression that we look for not only a good, but a *"good and fast"* packing or covering. In that paper we mainly dealt with only one possible *online* BP/CD model, and proposed a new method that we call the "Evolution of Algorithms". In case of such methods a neighborhood structure is defined among algorithms, and using a metaheuristic (for example simulated annealing) in some sense the best algorithm is chosen to solve the problem. Now we turn to investigate the *offline* case. We define several ways to treat such a BP/CD problem, although we investigate only one of them here. For the analysis, a novel view on "offline optimum" is introduced, which appears to be relevant concerning all problems where a final solution is ordering-dependent. We prove that if the item sizes are not allowed to be arbitrarily close to zero, then an optimal offline solution can be found in polynomial time. On the other hand, for unrestricted problem instances, no polynomial-time algorithm can achieve an approximation ratio better than 6/7 if $P \neq NP$.

[*]Department of Mathematics, University of Pannonia, Veszprém, Hungary, e-mail: dosagy@almos.vein.hu.

[†]Alfréd Rényi Institute of Mathematics, Hungarian Academy of Sciences, Budapest, Hungary, and Department of Computer Science and Systems Technology, University of Pannonia, Hungary. Email: tuza@dcs.vein.hu. Supported in part by the Hungarian Scientific Research Fund, OTKA grant T-81493.

# 1 Introduction

We deal with a new problem that we call *Bin Packing/Covering with Delivery*, BP/CD for short. In this kind of problems items are packed into bins, and the closed bins (into which we do not want to pack further items) are delivered. The objective is concerned with the delivered bins. For example, if we deal with bin covering with delivery, then some profit is paid for the covered and delivered bins, and this profit is to be maximized.

In case of the pure bin packing problem each item has a size, each item must be packed into some bin, the total size of the items being packed into a bin cannot exceed the capacity of the bin (it is generally normalized to 1), and the number of used bins is to be minimized. In case of bin covering a bin is *covered* if the total size of items packed into the bin is not less than 1, and we want to cover as many bins as possible. It is well known that both problems are NP-hard to solve optimally [6]. In many situations, however, one needs to find a feasible solution within a reasonable guaranteed time bound.

In the online setting, if the items arrive one by one, and an item needs to be packed immediately when it is revealed, the items are supposed to be very quickly packed; the "cost" of this speed is the decrease of efficiency with respect to the value of the solution. An example is the online packing algorithm called Next Fit, which uses only one open bin at a time; i.e., the next item is always packed into the actual opened bin if it fits, otherwise the bin is closed and a new one is opened to accommodate the new item. For this algorithm the number of used bins can be (in the worst case) the double of what it could be in an optimal solution.

We will deal with the offline setting in this paper, but it is still assumed that the items are ordered into a list $L$.

The objective depends not only on the *quality* of the packing or covering, but it is also influenced by the *speed* of the packing/covering procedure; i.e., the objective also depends on the current time of closing (and delivering) the bins. In this way we can handle also the time needed for making the packing/covering, so we are looking for a *"good and fast"* packing or covering.

There are several ways to define such a problem; in the next section we give some interesting versions that have never been investigated before. Then we will choose only one option, namely Problem 6 from the proposed versions: in the case under investigation it will be punished in the objective if "too many" bins are opened, since it is a natural idea that if more bins are open simultaneously then they need more time to handle.

There exist some works on "Scheduling with delivery", see [8] and the papers cited there, which is a similar problem; but bin packing or covering with delivery has never been treated before, at least up the the best knowledge of the authors.

In Section 2 we describe some problem variants which seem to be interesting but will not be discussed here in detail. Then in Section 3 we give the exact definition of the (offline) problem which is treated in this paper, and we also observe some properties of the problem. In Section 4 some efficient algorithms are proposed to determine the optimum for classes of problem instances where the item sizes are bounded from below by a positive constant. On the other hand, for unrestricted instances, no polynomial-time algorithm can achieve an approximation ratio better than $6/7$ if $P \neq NP$.

The online setting of the problem is not treated in this paper, since it was already investigated in [2], where as a main contribution, we defined a meta-heuristic that we call the *Evolution of Algorithms* (EOA for short). This concept can be viewed as the "online version" of Evolutionary Algorithms, which were proposed earlier in the literature to solve offline problems.

## 2 Some problems of interest

There are many possibilities to define such a problem: "Bin Packing/Covering with Delivery". Let us see some of them below.

**Problem 1.** This is an online problem, the items arrive according to a list $L$. At any time $t$, when an item is revealed (and we get to know the size of the item), we are also informed about the actual temporary delivery cost $c_t$, which also changes in time.

For simplicity we suppose that the revealing time of the $t$-th item is just $t$, i.e. the index of the item. Whenever a new item comes, we must immediately pack it either into an existing bin where it fits or into a newly opened bin, and then we are also allowed to close some bins if we want. If we close a bin just after the revealing of the $t$-th item, we deliver the closed bin, and we pay $c_t$ cost for the bin just delivered. We can close more bins at a time, as well. It is also allowed that no bin is closed: only put the actual item into a bin where it fits, and keep all opened bins open (not to deliver any bin at this moment yet, hoping that delivery cost will be decrease in the future). When the input sequence ends, all opened bins must be delivered at the actual (final) cost. The objective is to minimize the total delivery cost. Note that in case of $c_t = 1$ for all $t$, we get back to the pure online bin packing problem, thus this version is a generalization of it.

To make the problem reasonably determined, one may put some assumptions on the cost function, for example that (a) $\alpha \leq c_t \leq \beta$ holds for some fixed positive constants $\alpha, \beta$ for all $t$, or (b) the numbers $c_t$ form an increasing sequence.

**Problem 2**. A similar covering problem can be defined in the following alternative way. The items arrive one by one. At any time $t$, when item $I_t$ is revealed, we are also informed about the actual temporary profit $c_t$ which is gained for delivering (and selling the content of) the bin, if the bin is covered. Only the covered bins are allowed to be delivered. The objective is to maximize the total profit gained for the delivered bins. (If there remain some uncovered bins at the end of the sequence, no profit is received for them.) Again, this problem is a generalization of the pure online bin covering problem if $c_t = 1$ holds for all $t$.

**Problem 3**. We get a special but still interesting version of the packing Problem 1 if the cost/profit function is linear, $c(t) = At + B$ where $B \geq 0$. Then the change of the cost/profit function in the future becomes predictable.

For example, if $A = 0$ and $B = 1$, we get back the pure online problem. If $B = 1$ still holds, but $A > 0$ is a "small" real number, it means that the delivery cost of the bins (in the packing version) is slightly increasing. Then we are confronted with a trade-off between waiting with closing the bins to get better packings (but then the delivery cost will be bigger), or we can hurry to close the bins to keep the delivery cost low, but then we can not pack too well the items. Another interesting version is if the function $c(t)$ is piece-wise constant, with a few switches only. For example, we pay $C_1$ cost (or we get $C_1$ profit) if the bin is closed and delivered before some prefixed time $T$, and we pay $C_2 > C_1$ cost (or we get only $C_2 < C_1$ profit) after this prefixed time.

**Problem 4.** The next one is a covering problem. We gain money for each covered bin, but the time needed to cover it (i.e., the time elapsed between opening and closing the bin) is punished in the following way: A monotone increasing punishment function $G : \mathbb{N} \to \mathbb{R}^+$ is given. If the minimum and maximum indices of items packed into a covered bin are $i$ and $j$, respectively, then the value of punishment is $G(j - i)$. The objective is to maximize the number of covered bins minus the total punishment. (In case if $G = 0$ is the identically zero function, we get back again the pure online bin covering problem.)

A packing problem of similar flavor can be defined as follows. The objective is to minimize the number of used bins (i.e., where we pack the items) plus the total punishment which is defined for each bin as before. This kind of problem is proposed in [5].

**Problem 5.** This is another covering problem. The items arrive again in an online fashion, but we have a buffer that can accommodate at most $K$ items at any time. If the buffer is full and the next item arrives, one of the $K+1$ items must immediately be packed into some bin; but any item may be stored in the buffer for an arbitrarily long time. At the end of the procedure, when no more items come, each item being in the buffer must also be packed. Delayed items are punished only when we close and deliver a bin; we then must pay $G(k) \geq 0$ cost ($0 \leq G(k) < 1$) if the buffer contains exactly $k$ items at the moment of delivery. In this way, upon delivery we gain $1 - G(k)$ profit. The goal is to maximize the total profit. (By assumption we have $k \leq K$; and $G(k)$ is supposed to be non-decreasing in $k$, to punish storing more items in the buffer, since it needs more work to handle more items. In case $G(k) = 0$ we get the online covering problem with allowing the use of a buffer of size $K$, with no cost for using the buffer.) We note that there are some works on packing or scheduling with the usage of a reordering buffer, see [1, 3, 4, 7].

In this paper we will study a sixth problem, to be introduced next.

## 3 Problem definition

Now we choose only one from the many options, a covering problem different from the ones above, and we will deal only with this problem in the rest of this paper. Investigating the other interesting/important variations remains the topic of future research. Our problem can be stated as follows. Items arrive one by one according to a list $L$, the $i$-th item has size $p_i > 0$, and infinitely many unit capacity bins are available. Just when a new item is revealed, it must be packed into a bin. The number of items is denoted by $n$; it is known in the offline case, but in the online fashion it is not known in advance, the value of $n$ is revealed only at the end when no more items come and all items have been packed. A fixed integer $K > 0$ is also given, and there can be at most $K$ opened bins simultaneously.

A covering algorithm has the option of packing the current item into an already opened bin, or if the number of open bins is less than $K$, a new bin can also be opened and then the new item must be packed into it.

A bin $B$ is covered if the sum of sizes already packed into $B$ is not less than 1. The objective is defined by a profit function $G : \{1, \ldots, K\} \to \mathbb{R}^+$ as follows: If there are $k$ opened bins ($1 \leq k \leq K$) when bin $B$ becomes covered (at this moment the bin is delivered), then $G(k)$ profit is realized for covering this bin. Naturally, if a bin is covered and delivered, the number of open bins decreases by one.

We suppose that $G$ is a positive, monotone non-increasing function. By this assumption we model that handling more opened bins we need more time to decide where to pack the actual item. The objective is to maximize the total profit received for the closed and delivered bins.

For a real-life application let the following one stand here. At a small producer, some kind of fruit is packed into small boxes, the fruit arrives through a little window, each piece of fruit must be packed by hand into a box, and there is only one person on the other side of the window who manages this procedure. It is natural to suppose that this person cannot handle too many open bins simultaneously. Modeling this situation it appears quite natural that he is required to do as good covering as possible but on the other hand keeping as few bins opened as possible. Since these two objectives contradict each other, we simulate his position with a decreasing amount of money which can be gained for a covered bin as the number of opened bins increases.

In case of offline problems, the "goodness" of an algorithm is usually measured by approximation ratio, which is defined as follows. The value $C_A(I)$ of the solution determined by an offline algorithm $A$ on an input sequence $I$ is compared to the offline optimum value $OPT(I)$. In case of a maximization problem (as in our case, too) the infimum of the ratio $C_A(I)/OPT(I)$ taken over all sequences $I$ is called the approximation ratio of algorithm $A$.

To apply this principle it is substantial to specify what should be meant under "offline problem" and "offline optimum" if the items are ordered in a list $L$.

In case of offline problems it is always assumed that all information is available in advance about the input, thus we also suppose that all item sizes $p_i$ are given. On the other hand, if the items could be packed into the bins in an arbitrary order, then the role of the profit function $G$ would be totally neglected.

Let us put here a parenthetical remark. Knowing everything about the instance, one can always compute an optimum solution for unrestricted bin packing (although usually in exponential time). Then knowing how the items must be packed optimally, the optimum packing procedure can be made in a sequential manner (but generally different from the given list $L$!). And then, at each moment, one needs to have only one opened bin. Therefore, with unlimited computational capacity, a kind of offline optimum is achieved as a multiple of $G(1)$, regardless of the values $G(k)$ for $k \geq 2$.

For this reason, we define the offline model in a somewhat different way, as follows. All information concerning the input is known in advance, but the items must be packed into the bins according to the given list $L$, in the order as the items really come.

Certainly, also such an offline optimum solution exist for every finite input. At any moment when a new item comes, the problem solver can choose from finitely many options (there are at most $K$ bins where the next item can be packed), and among the all-in-all finitely many options, there is a solution which produces the best value of the objective function. Of course, the offline optimum depends not only on the item-set, but also on the given list $L$, and on the profit function $G$ as well.

After these comments, for any finite list $L$ of the items to be packed, and for any profit function $G$, let $C_A(L, G)$ mean the value of the solution generated by an offline algorithm $A$. We compare it to $OPT(L, G)$, the offline optimum value respecting the order specified by $L$. Then the approximation ratio of $A$ is the supremum of real $\rho$ ($0 \leq \rho \leq 1$) for which

$$C_A(L, G)/OPT(L, G) \geq \rho$$

holds for all $L, G$.

## 4 Theoretical results

If the items are not ordered in a list, then to compute the value of the offline optimum is NP-hard in the strong sense in general, because it contains 3-PARTITION as a subproblem. Also, if no special assumptions are given for the sequence $L$ and/or for the profit function $G$, the problem is hard. Indeed, if $G$ is a constant function, we get the classical (NP-hard) Bin Covering problem; and if $G$ is not constant, to solve the problem seems to be even much harder.

First we show that the offline optimum can be determined efficiently if the item sizes are bounded by a positive value from below (from above they are bounded by the bin size), with an arbitrary choice of the profit function.

**Theorem 1** *Let $K \in \mathbb{N}$ and $b \in \mathbb{N}$ be fixed integers, $G : \{1, \ldots, K\} \to \mathbb{R}^+$ any profit function, and $c > 0$ a fixed real.*

(i) *If all items of the input are required to be at least $c$, then for any given $L$ the offline optimum can be determined in polynomial time.*

(ii) *If the items of each input sequence can have at most $b$ different sizes, none of them being smaller than $c > 0$, then for any given $L$ the offline optimum can be determined in linear time.*

**Proof.** Since at most $K$ bins are open in any moment during a feasible packing procedure, we can assign an integer between 1 and $K$ to any bin, under the condition that no two bins of the same label are open at any time. Once this assignment is given, each item can get the label of the bin into which we pack it. Hence, each feasible packing procedure determines a sequence in $\{1, \ldots, K\}^n$ and also conversely, every such sequence uniquely determines a packing procedure.

Note further that the value of a solution can be computed from its sequence. Indeed, one can detect the contents of the opened bins step by step, identify the moments when a bin gets covered, look at the number of opened bins at that moment, and finally compute the corresponding sum of values of $G$.

Moreover, in each step, the essential pieces of information in the current situation can be characterized by the sum of $G$-values over the previously covered bins, and the current contents of the opened bins. Our key observation will be that in each step the possible number of these characteristics is bounded above by a polynomial of $n$ if the condition of (i) holds, and even by a constant under the condition of (ii).

*Proof of (i).* Let $m = \lfloor 1/c \rfloor$. Then any $m+1$ items completely cover a bin, therefore at most $M = \sum_{i=1}^{m} \binom{n}{i}$ different subsets can be the contents of an opened and yet uncovered bin. Thus, the at most $K$ open bins can have no more than

$$M^* := \sum_{i=0}^{K} \binom{M}{i}\binom{K}{i} i! = O(n^{K/c})$$

distributions of subsets as their contents in each step, where the factor $\binom{K}{i}i!$ stands for the possible label distributions ranging between 1 and $K$.

Based on these observations we design an efficient algorithm, which has the flavor of dynamic programming. For $t = 1, 2, \ldots, n$ we compute a list of all possible contents of the open bins, together with the maximum achievable profit belonging to those contents. As shown above, for each $t$ there are at most $M^*$ partial solutions to be listed for the subsequence $p_1, \ldots, p_t$. In principle the next item $p_{t+1}$ can be packed into any of the at most $K$ bins, this yields $KM^*$ (or fewer) candidates for a partial solution on $p_1, \ldots, p_{t+1}$. Some of them may coincide; we keep the best $G$-value for each distribution. In this step we also keep track of the former distribution from which the best $G$-value has been obtained.

Once all distributions have been determined for $p_n$, the offline optimum is achieved by one having the largest $G$-value. The algorithm runs in $O(n^{1+K/c})$ time.

*Proof of (ii).* Let again $m = \lfloor 1/c \rfloor$, which is an upper bound on the number of items in an opened but not yet delivered bin. If an open bin contains exactly $j$ items ($1 \leq j \leq K$), its current load can be one of at most $\binom{b-1+j}{j}$ different combinations from the possible $b$ different values. Hence, the load of an open bin cannot take more than

$$\sum_{j=1}^{K} \binom{b-1+j}{j} < \binom{b+K}{K}$$

values between 0 and 1. With $i$ labeled open bins ($i = 1, \ldots, K$) this means altogether fewer than

$$\sum_{i=1}^{K} \binom{b+K}{K}^i \binom{K}{i} i!$$

different situations for any $t$. Hence, the algorithm described above requires at each step to check a bounded number of load distributions only. Consequently, the best $G$-value for $p_n$ can be determined in $O(n)$ time. Then, in $n$ steps we can trace back a sequence in $\{1, \ldots, K\}^n$ that achieves optimum. This sequence describes an optimal packing procedure. □

Now we prove that there is no APTAS (asymptotical polynomial time approximation scheme) for the offline problem. More exactly, we show the following result:

**Theorem 2** *For a suitable choice of the profit function $G$, there exists a class of item-sequences $L$ for which no polynomial-time algorithm can achieve an approximation ratio (neither absolute nor asymptotic) better than 6/7, for any integer $K \geq 2$, if $P \neq NP$.*

**Proof.** We prove that the following choice is suitable, the last part applying only if $K \geq 3$:

$$G(1) = 1, \qquad G(2) = 1/2, \qquad G(3) = \ldots = G(K) = 0.$$

We describe a generic class of item-sequences, which are composed of the following subsequences. First, in phase (a) many small items with total size 2 arrive, followed in phase (b) by two big items, both with size 1. We say that this is one batch in the sequence. The sequence consists of $n$ batches, each batch consist of the same items in the same order. We suppose that there exists a partition of the many small items into two equal sums, but no starting part of the sequence of the small items has total size exactly 1. For the (a) phases this means the hard-to-recognize class of *positive instances* (i.e., those admitting an optimal solution) for the well-known PARTITION problem, normalized to a fixed sum.

Our first claim is that $OPT \geq n \cdot (3G(1) + G(2)) = 3.5\,n$. Indeed, let us consider the following offline solution. At the beginning two bins are opened and the small items are packed into these two bins in such a way that at the end of phase (a) both bins contain items with total size 1, i.e. both bins are covered. Then we gain $G(2) = 0.5$ profit for the bin covered first and $G(1) = 1$ for the second one. At this point there are no open bins. Then the two big items come one by one, we put them into one open bin and again one open bin, and we gain $2\,G(1) = 2$ profit for covering these bins. The sequence consists of $n$ batches, and in each batch we gain $3\,G(1) + G(2)$ profit. (This is actually the optimal offline solution of the problem for the given input, but we do not need this fact in the proof.)

On the other hand, let us consider an arbitrary polynomial-time algorithm $A$. We describe the proof first under the assumption that $A$ never opens more that two bins.

Since the PARTITION problem is NP-hard to solve, $A$ cannot cover two bins with the small items of any batch in a generic instance. Thus, if there are no opened bins at the beginning of a phase (a), then at least one bin remains opened at the end of that phase. Therefore, qualifying the situations according to the number of opened non-empty bins, the following transients are possible:

| started | finished | maximum profit |
|---|---|---|
| 0 | 1 | $G(1)$ |
| 0 | 2 | $G(1)$ |
| 1 | 0 | $2\,G(1)$ |
| 1 | 1 | $2\,G(1)$ |
| 1 | 2 | $2\,G(1)$ |
| 2 | 0 | $2G(1)+G(2)$ |
| 2 | 1 | $2G(1)+G(2)$ |
| 2 | 2 | $2G(1)+G(2)$ |

The possible transients after Phase (a)

Those values are easily checked, for example if phase (a) starts with two opened bins then the first cover has profit $G(2)$ only. A similar table can be composed for phase (b), too:

| started | finished | maximum profit |
|---|---|---|
| 0 | 0 | $2\,G(1)$ |
| 1 | 0 | $2\,G(1)$ |
| 1 | 1 | $2\,G(2)$ |
| 2 | 0 | $G(1)+G(2)$ |
| 2 | 1 | $2\,G(2)$ |

The possible transients after Phase (b)

To verify this, one should observe that an item of size 1 always covers a bin, hence at most one bin can remain open and the end of the phase; and if it is indeed one, then both large items cover a second bin.

The composition of those two tables describes the possible kinds of transition during one batch. Note that the process begins with no items (i.e., no bins can have nonzero contents), hence under the assumption that a third bin never is open, at most one bin remains opened at the end of each batch. Thus, we obtain:

| started | middle | finished | maximum profit |
|---|---|---|---|
| 0 | 1 | 0 | $3\,G(1)=3$ |
| 0 | 2 | 0 | $2\,G(1)+G(2)=2.5$ |
| 0 | 1 | 1 | $G(1)+2\,G(2)=2$ |
| 0 | 2 | 1 | $G(1)+2\,G(2)=2$ |
| 1 | 0 | 0 | $4\,G(1)=4$ |
| 1 | 1 | 0 | $4\,G(1)=4$ |
| 1 | 2 | 0 | $3\,G(1)+G(2)=3.5$ |
| 1 | 1 | 1 | $2\,G(1)+2\,G(2)=3$ |
| 1 | 2 | 1 | $2\,G(1)+2\,G(2)=3$ |

One batch with intermediate state between phases (a) and (b)

The main point here is the maximum achievable profit during a batch. This means four possible combinations for each batch (starting with 0 or 1 and ending with 0 or 1), except that we have only two cases for the first batch (starting with 0). Some combinations (namely, 0–2–0 and 1–2–0) are irrelevant because they are inferior to their alternatives. In this way the four possible transients belonging to the elements of $\{0,1\} \times \{0,1\}$ have respective profits $3, 2, 4, 3$.

For precise analysis, the possibilities for the process of packing the batch sequence will be conveniently represented with an edge-weighted directed graph $G = (V, E)$ having $2n + 1$ vertices and $4n - 2$ edges, where

$$V = \bigcup_{i=0}^{n} V_i, \qquad V_0 = \{v_{0,0}\}, \qquad V_i = \{v_{i,0}, v_{i,1}\} \quad \text{for} \quad i = 1, \ldots, n,$$

$V_{i-1}$ and $V_i$ are completely joined, oriented from smaller index to larger, and the edge weights are defined as

$$w(v_{i-1,0}v_{i,0}) = 3 \qquad (1 \le i \le n),$$
$$w(v_{i-1,0}v_{i,1}) = 2 \qquad (1 \le i \le n),$$
$$w(v_{i-1,1}v_{i,0}) = 4 \qquad (2 \le i \le n),$$
$$w(v_{i-1,1}v_{i,1}) = 3 \qquad (2 \le i \le n).$$

Now, $C_A(L, G)$ cannot be larger than the maximum weight of a directed path from $V_0$ to $V_n$. Let us denote this maximum by $\ell$. We claim that

$$\ell = 3n$$

from what we shall conclude

$$\frac{C_A(L, G)}{OPT} \le \frac{\ell}{OPT} \le \frac{3n}{3.5n} = \frac{6}{7}.$$

In order to prove $\ell = 3n$, let $P$ be a $V_0$–$V_n$ path of maximum weight, such that the number of vertices of type $v_{i,0}$ is as large as possible. We claim that the vertex set of $P$ is just $\{v_{i,0} \mid 0 \le i \le n\}$, from what the assertion will follow immediately.

Let us observe that $v_{n,0}$ surely belongs to $P$. This fact is valid because $w(v_{n-1,0}v_{n,0}) > w(v_{n-1,0}v_{n,1})$ and $w(v_{n-1,1}v_{n,0}) > w(v_{n-1,1}v_{n,1})$. Hence, let us suppose for a contradiction that $v_{i,0} \notin P$ for some $0 < i < n$, and let $i$ be the largest subscript with this property. Then of course $v_{i+1,0} \in P$.

Let $j \in \{0, 1\}$ be the value for which $v_{i-1,j} \in P$. Now we can observe that
$$w(v_{i-1,j}v_{i,0}) + w(v_{i,0}v_{i+1,0}) = w(v_{i-1,j}v_{i,1}) + w(v_{i,1}v_{i+1,0})$$
independently of the actual value of $j$. Therefore, replacing $v_{i,1}$ with $v_{i,0}$ in $P$ does not decrease the total weight, moreover it increases the number of vertices with second subscript 0. This contradicts the choice of $P$, and hence proves $\ell = 3n$.

To complete the proof of the theorem, we need to verify that three or more open bins do not help. The argument is an extended version of the one above. The upper bound on the possible profit in phase (a) started with more than two open bins is $2G(2) + G(1)$, the same as starting with two open bins, independently of the number of bins left open after this phase. Also, if the number of open bins at the beginning is 0 or 1, it is irrelevant for the calculation of profit whether we close phase (a) with at most two or at least three open bins, i.e. the data in the earlier table remain valid for the current cases, too.

The difference for phase (b) is that more than one bin may remain open afterwards, and this allows us to start the next batch with more than one bin, yielding a higher profit in this batch. It will turn out, however, that doing so has a high price in the preceding bin and after all we do not get any extra profit. In any case, each of the two large items of phase (b) covers the bin where it is packed, hence the number of opened bins cannot increase. The maximum profit theoretically achievable in the cases not treated before is listed in the following table.

| started | middle | finished | maximum profit |
|---|---|---|---|
| $\geq 2$ | 0 | 0 | $4\,G(1) + G(2) = 4.5$ |
| $\geq 2$ | 1 | 0 | $4\,G(1) + G(2) = 4.5$ |
| $\geq 2$ | 1 | 1 | $2\,G(1) + 3\,G(2) = 3.5$ |
| $\geq 2$ | 2 | 0 | $3\,G(1) + 2\,G(2) = 4$ |
| $\geq 2$ | 2 | 1 | $2\,G(1) + 3\,G(2) = 3.5$ |
| $\geq 2$ | 2 | 2 | $2\,G(1) + G(2) = 2.5$ |
| $\geq 2$ | 3 | 1 | $2\,G(1) + 2\,G(2) = 3$ |
| $\geq 2$ | 3 | $\geq 2$ | $2\,G(1) + G(2) = 2.5$ |
| $\geq 2$ | $\geq 4$ | $\geq 2$ | $2\,G(1) + G(2) = 2.5$ |

Upper bound on profit in a batch starting with at least two open bins

As one can observe, the respective upper bounds 4 / 3 / 2 for one starting and 0 / 1 / 2 finishing open bins get increased to 4.5 / 3.5 / 2.5 for two starting bins. The extra profit is 0.5 while the loss in the preceding batch is at least 1 to finish with more than 1 open bin.

A precise formal analysis can be done by extending the digraph considered previously: vertices and edges represent the possible numbers of open bins before/after each batch and the transitions between them, respectively; and edge weights correspond to the profits of the transitions. The digraph can be reduced to a relevant smaller subdigraph by the following observations:

- decreasing the number of starting bins from $> 2$ to 2 keeps the bound on profit unchanged;

- decreasing the number of finishing bins from $> 2$ to 2 does not decrease the bound on profit;

- decreasing the number of middle bins from $> 2$ to 2 does not decrease the bound on profit.

Hence, optimal paths need not use any parts of the digraph other than levels $0, 1, 2$. Finally, any directed path containing vertices on level 2 can be modified to another path that has one fewer vertices on level 2, without decreasing the upper bound on total profit. This implies by repeated application that more than two open bins do not help, hence completing the proof of the theorem. □

We note that there cannot be a PTAS in the absolute sense either. This statement is easy to verify, an instance class with absolute lower bound 2/3 on the performance ratio can be derived from bad (non-partitionable) instances of the partition problem, which is NP-hard to solve.

## 5 Conclusions

In this paper we proposed a new problem, which is quite natural but never treated before, we call it "Bin Packing/Covering with Delivery". This mainly means that we are looking for not only a good packing or covering, but a "good and fast" one. This approach seems to be interesting also in a more general setting: for any (hard) algorithmic problem it can be of great importance to obtain a high-quality solution in a time-efficient way.

Our present offline setting substantially differs from standard bin packing/covering in the way that the items have to be packed in a prescribed order. This condition changes the structure of problem considerably. In addition, the number of open bins is taken into account, too.

We have investigated the offline scenario from a theoretical point of view. On the positive side, efficient algorithms have been proposed which solve the problem in polynomial time, or even in linear time, under some assumptions (if no items are "too small", or if the number of different item sizes is bounded).

On the negative side, a non-approximability result has been proved. Namely, if $P \neq NP$, then a class of problem instances can be defined such that the best offline solution computable in polynomial time for a worst-case generic instance from this class is at most $6/7$ times the optimum. On the other hand, it is easy to give a (linear-time) algorithm that gives at least half of the optimum value for any list of items and for any gain-function: Dual Next Fit is such an algorithm, as it is proved in [2]. Hence, there remains a gap between $1/2$ and $6/7$, leaving the open problem to determine the best ratio achievable by a polynomial-time algorithm for all item lists and all gain-functions.

Finally we note that similar investigations could also be done concerning the other five problems, which are defined in Section 2.